\documentclass[jcp, floatfix, nobibnotes, reprint, superscriptaddress]{revtex4-1}
\usepackage{docs}%
\usepackage{siunitx}
\usepackage{amsmath}
\usepackage{booktabs}
\usepackage{amssymb}
\DeclareSIUnit[number-unit-product = {\,}]
\cal{cal}
\DeclareSIUnit\kcal{\kilo\cal}
\DeclareSIUnit[number-unit-product = {\,}]
\Btu{Btu}
\DeclareSIUnit[number-unit-product = {\,}]
\Fahr{\degree F}
\DeclareSIUnit[number-unit-product = {\,}]
\usepackage{bm}%

\usepackage{graphicx}
\usepackage[utf8]{inputenc}
\usepackage{tabularx} 

\usepackage{dcolumn}
\usepackage{epstopdf}
\usepackage{afterpage}
\usepackage{setspace}
\usepackage{multirow}
\usepackage{dsfont}
\usepackage{sidecap}
\usepackage{epstopdf}
\usepackage{braket}
\usepackage{siunitx} 
\usepackage{color, colortbl}

\definecolor{Gray}{gray}{0.9}

\AppendGraphicsExtensions{.gif}

\DeclareMathAlphabet\mathbfcal{OMS}{cmsy}{b}{n}

\setcitestyle{super}

\begin{document}
\setstretch{1.0}

\title[]{Generalized convolutional many body distribution functional representations}

\author{Danish Khan}

\affiliation{Chemical Physics Theory Group, Department of Chemistry, University of Toronto,
St. George Campus, Toronto, ON, Canada}
\affiliation{Vector Institute for Artificial Intelligence, Toronto, ON, M5S 1M1, Canada}

\author{O. Anatole von Lilienfeld}
\email{anatole.vonlilienfeld@utoronto.ca}
\affiliation{Chemical Physics Theory Group, Department of Chemistry, University of Toronto, St. George Campus, Toronto, ON, Canada}
\affiliation{Department of Materials Science and Engineering, University of Toronto, St. George Campus, Toronto, ON, Canada}
\affiliation{Vector Institute for Artificial Intelligence, Toronto, ON, M5S 1M1, Canada}
\affiliation{ML Group, Technische Universit\"at Berlin and Institute for the Foundations of Learning and Data, 10587 Berlin, Germany}
\affiliation{Berlin Institute for the Foundations of Learning and Data, 10587 Berlin, Germany}
\affiliation{Department of Physics, University of Toronto, St. George Campus, Toronto, ON, Canada}
\affiliation{Acceleration Consortium, University of Toronto, Toronto, ON}

\begin{abstract}
Modern machine learning (ML) models of chemical and materials systems with billions of parameters require vast training datasets and considerable computational efforts. Lightweight kernel or decision tree based methods, however, can be rapidly trained, leading to a considerably lower carbon footprint. We introduce generalized convolutional many-body distribution functionals (cMBDF) as highly compute and data efficient atomic representations for accurate kernels that excel in low-data regimes. Generalizing the MBDF framework, cMBDF encodes local chemical environments in a compact fashion using translationally and rotationally invariant functionals of smooth atom centered Gaussian electron density proxy distributions weighted by interaction potentials. The functional values can be efficiently evaluated by expressing them in terms of convolutions which are calculated via fast Fourier transforms and stored on pre-defined grids. In the generalized form each atomic environment is described using a set of functionals uniformly defined by three integers; many-body, derivative, weighting orders. Irrespective of size/composition, cMBDF atomic vectors remain compact and constant in size for a fixed choice of these orders controlling the structural and compositional resolution. While being up to two orders of magnitude more compact than other popular representations, cMBDF is shown to be more accurate for the learning of various quantum properties such as energies, dipole moments, homo-lumo gaps, heat-capacity, polarizability, optimal exact-exchange admixtures and basis-set scaling factors. Applicability for organic and inorganic chemistry is tested as represented by the QM7b, QM9 and VQM24 data sets. Due to its compactness, model training and testing times are reduced from 23 hours to 8 minutes, implying a corresponding reduction in carbon footprint.
\end{abstract}

\maketitle
\section{Introduction}

\begin{figure*}[!htb]
    \centering
    \includegraphics[width=\linewidth]{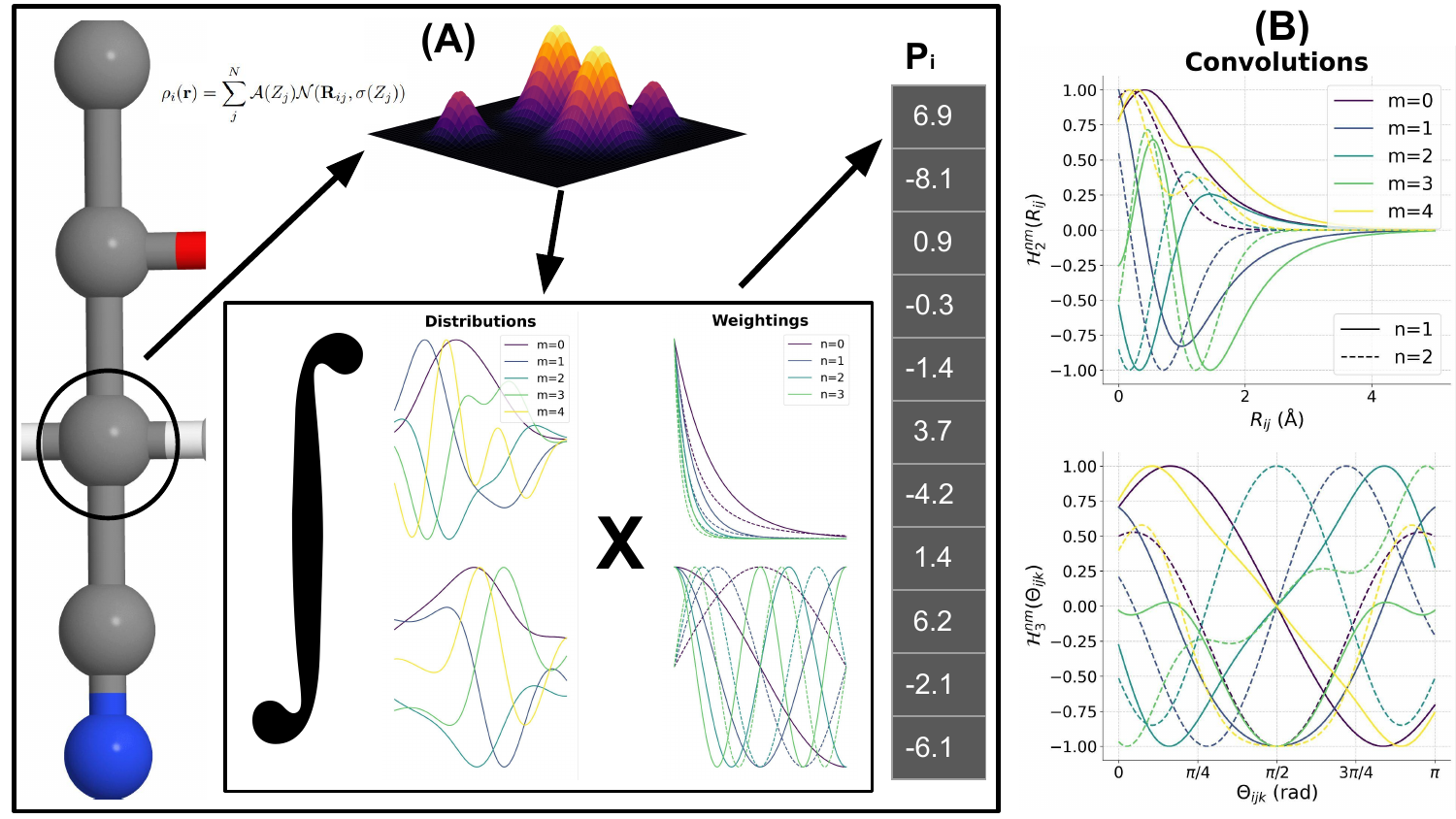}
    \caption{(A) Schematic representation of cMBDF atomic feature vector generation for an atom $i$ within any system.
    A smooth atomic density $\rho_{i}(\mathbf{r})$ (eq.~\ref{rho}), centered on atom $i$, is constructed by placing element specific basis functions on all atoms within its local neighbourhood. 
    Translationally and rotationally invariant distributions based on this atomic density are generated by projecting it onto many-body internal coordinates (eqs.~\ref{rho2},\ref{rho3},\ref{rho4_1}).
    The distributions and their derivatives are weighted by many-body interaction potentials of interest (eqs. ~\ref{gn2},~\ref{gn3}).
    The chemical environment of the atomic species is then encoded in a feature vector consisting of these functional values integrated over the domain of the distributions (eqs.~\ref{p2nm},~\ref{p3nm_final},~\ref{p4nm_final}).
    The integral evaluations can be bypassed for "on-the-fly" generation of the feature vectors by expressing them as sums of convolutions (eqs.~\ref{h2nm},~\ref{h3nm},~\ref{h4nm}). 
    Figure (B) shows the two (left) and three-body (right) convolved functions $\mathcal{H}_{\nu}^{nm}$ for weighting functions of the first type in eqs.~\ref{gn2},~\ref{gn3}.} 
    \label{fig:convs}
\end{figure*}
While training data needs are increasingly being met by high quality quantum mechanical (QM) datasets~\cite{qm9,ANI-1,QMugs,qm7b,vqm24}, it is crucial to recognize that chemical space is considerably larger and that, being fundamentally interpolative, ML models inherently lack the universal applicability of the Schrödinger equation.~\cite{li2023critical}.
This is particularly concerning as training large deep learning models already consume significant energy and are on a clear track for becoming increasingly unsustainable~\cite{strubell-etal-2019-energy}.
Carbon dioxide emissions from training large language models, for instance, are soon expected to match the monthly emissions of New York City~\cite{DL_carbonf}.
Consequently, navigation of chemical compound space solely using general purpose ML models is impractical due to their inherent extrapolation limitations~\cite{li2023critical}.
\\
Alternatively, they can be paired with extrapolative quantum chemistry methods through $\Delta$~\cite{deltaML}, multi-level~\cite{m3l}, and adaptive~\cite{apbe0, abasis} learning schemes which lead to significant reduction in training requirements.
Substantial reductions, with or without the aforementioned schemes, are also obtained through improved atomic feature vector mappings~\cite{communication_bing,wigner_kernels} (or representations) by incorporating known physical laws.
Some of the most prominent examples of these are the Behler-Parrinello symmetry functions~\cite{Behler-Parrinello_NN,acsf} (ACSF), permutationally invariant polynomials~\cite{pip}, Coulomb matrix~\cite{CM} (CM), smooth overlap of atomic positions~\cite{soap} (SOAP) and atomic cluster expansion~\cite{ace}.
While such physics-based representations can significantly reduce training data needs~\cite{physics-inspired-reps-ceriotti}, the most recent deep learning based methods rely on feature learning through the data itself which bloats the training requirements.
The data-efficiency is especially pronounced when using these "hand-crafted" representations with more efficient interpolants in the low training data regime such as kernel-based methods~\cite{gary2023}.
Furthermore, the computational cost of these models is significantly affected by the choice of the molecular representation~\cite{mbdf}.
\\
Due to the profound impact on all learning tasks and computational cost, the atomic representation choice is akin to the level-of-theory in Pople's model quantum chemistry methods.
Expectedly then, the use of physically inspired 
representations leads to more data efficient ML methods which do not rely on first learning the mapping through the training data itself.~\cite{wigner_kernels}
This is crucial due to the vast size of chemical space which can only be probed through interpolative models if they (i) can be regressed with limited training data, (ii) can be applied across structural and compositional degrees of freedom and (iii) are computationally feasible enough (including training time) to provide significant acceleration over their, extrapolative, quantum chemistry counterparts.
\\
Satisfying these requirements while striking the optimal trade-off between data and computational efficiency (see below), in this work we introduce the convolutional many body distribution functionals (cMBDF) representation.
Generalizing the MBDF framework~\cite{mbdf}, we use a uniform series of translationally and rotationally invariant functionals of the atomic density to efficiently quantify the local chemical environment of an atom.
The original MBDF representation focused on compactness and used only 5 functionals, chosen empirically to maximize accuracy, to describe each atomic environment.
This framework is generalized to obtain a systematically improvable family of atomic descriptors controlled by three integers (weighting function, many-body and derivative orders) which allow controlling the computational vs data efficiency trade-off using a set of uniformly defined functionals.
For a fixed value of these indices; the atomic representation remains constant size due to its invariance to the radial cut-offs, number of neighbours and unique chemical elements.
Apart from element specific basis functions, improved weighting functions and four-body functionals, cMBDF bypasses all integral evaluations (performed numerically in MBDF) by expressing the functionals as a series of convolutions. 
Using the convolution theorem, these are efficiently evaluated via fast Fourier transforms and stored on pre-defined grids leading to significant speed-ups for "on-the-fly" application and gradient evaluation. 
Physical interactions (both short and long-range) of interest can be incorporated by raising/lowering the weighting function order.
\\
The compactness of cMBDF feature vectors is inherent to the methodology in contrast to compression techniques~\cite{Darby2022} which can be applied to all atomic representations.
Hence, it is able to outperform other commonly used representations while remaining upto 2 orders of magnitude more compact as demonstrated for several learning tasks below.
This computational and data efficiency has allowed its successful application to adaptive-ML schemes~\cite{apbe0,abasis} across chemical space which improve existing quantum chemistry methods with limited, high quality, training data. 

\section{Theory and methods}
\begin{figure*}[!htb]
    \centering
    \includegraphics[width=\linewidth]{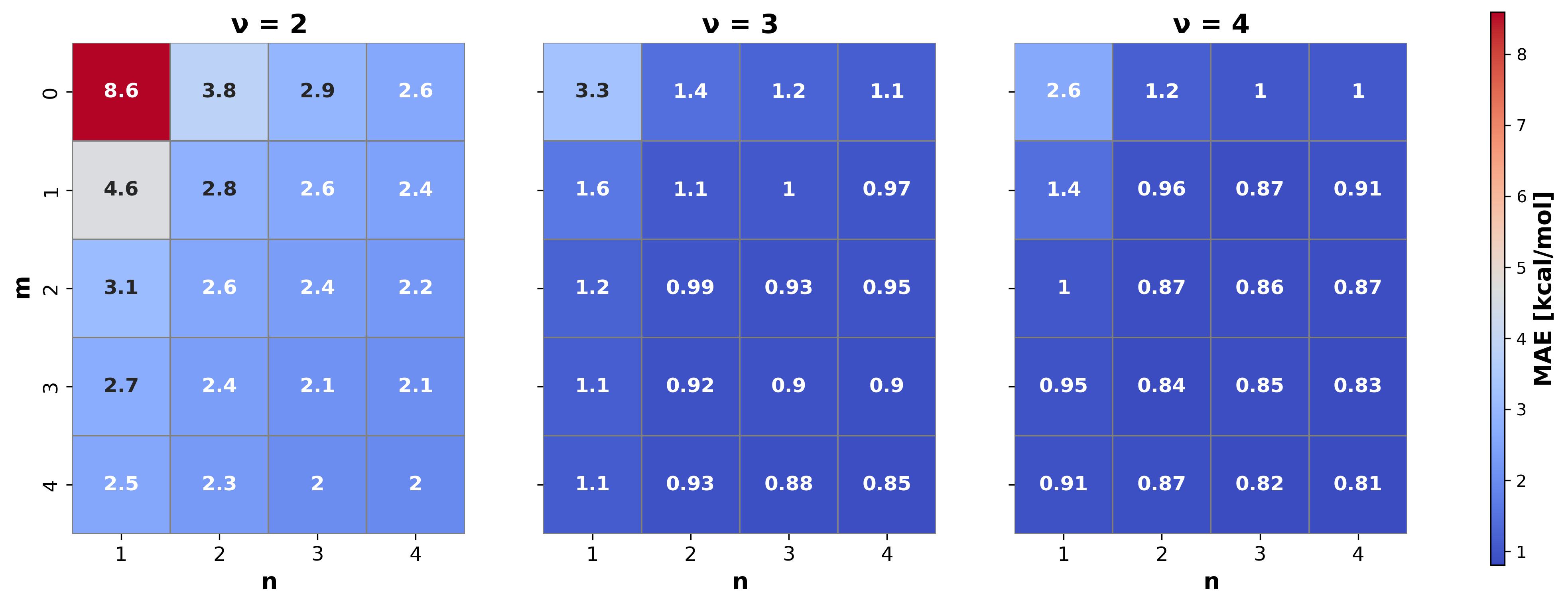}
    \caption{Heat-maps showing variation in predictive accuracy with the three integers $\nu$ (many-body order), $m$ (derivative order), $n$ (weighting function order) controlling the functionals generated by cMBDF. 
    Mean absolute errors (MAE) for prediction of atomization energies 
    using a kernel ridge regression (KRR) model are shown from the QM7b~\cite{qm7b} dataset.
    A training/testing-split size of 2000/1000 was used.} 
    \label{fig:errors_gs}
\end{figure*}
\subsection{Two and three-body functionals}
We are interested in defining a feature vector mapping ${\bf P}_{i}$ of an atom $i$ within a chemical system while minimizing the vector length.
The starting point is the smooth atom-centered atomic density~\cite{physics-inspired-reps-ceriotti} (an electron density proxy) $\rho_{i}(\mathbf{r})$
\begin{equation}
    \rho_{i}(\mathbf{r}) = \sum_{j}^{N} \mathcal{A}(Z_{j}) \mathcal{N}(\mathbf{R}_{ij}, \sigma(Z_{j}))
    \label{rho}
\end{equation}
where $\mathcal{N}(\mu, \sigma) = \frac{1}{\sqrt{2 \pi \sigma^2}}\exp({-\frac{(x-\mu)}{2\sigma^2}})$ denotes the normal probability distribution function (PDF) centered at $\mu$ with standard deviation $\sigma$ throughout this work, $\mathbf{R}_{ij} = \mathbf{R}_{j} - \mathbf{R}_{i}$ is the relative position of atom $j$ with respect to $i$ and $N$ denotes the number of atoms within a local cut-off radius $r_{\rm cut}$ around $i$.
The standard deviation $\sigma(Z_{j})$ of each basis function corresponds to the van-der-Waals radius ($r_{\rm vdW}$) of the element with nuclear charge $Z_{j}$ and the scaling pre-factor $\mathcal{A}_{2}(Z_{j})$ encodes the chemical identity of atom $j$ in terms of group and period (see eq.~\ref{alc_scale}).
Incorporating rotational invariance, the density can be projected onto internal coordinates starting with the 2-body distribution function $\rho_{i}(r)$
\begin{align}
    \rho_{i}(r) &= \braket{{\rho_{i}(\mathbf{r})}|\sum_{j}\delta(R_{ij} - r)|{\rho_{i}(\mathbf{r'})}}
\end{align}
which can be simplified using the fact that the convolution of two Gaussians is another Gaussian
\begin{align}
    \rho_{i}(r) &\approx \sum_{j}^{N} \mathcal{A}_{2}(Z_{j}) ~\mathcal{N}(R_{ij}, \sigma(Z_{j}))
    \label{rho2}
\end{align}
where $R_{ij} = |\mathbf{R}_{j} - \mathbf{R}_{i}|$ is the inter-atomic distance between atoms $i$ and $j$.
Translationally and rotationally invariant feature vector components encoding 2-body features of the atomic environment can then be defined through functionals of the general form
\begin{align}
    P_{2}^{nm}[i] =\int_{0}^{\infty}dr~g_{n2}(r) ~\partial_{r}^{m}\rho_{i}(r)
    \label{p2nm}
\end{align}
where $g_{n2}(r)$ are suitable weighting functions.
Inclusion of the derivatives $\partial_{r}^{m}\rho_{i}(r)$ allows unique description of the {\em total} $\rho_{i}(r)$ in eq.~\eqref{rho2} without generating separate distributions for each chemical element in the neighbourhood.
Eq. ~\eqref{p2nm} can be expressed in terms of $m$-th degree Hermite polynomials $H_{m}$ centered at $R_{ij}$
\begin{widetext}
\begin{align}
    P_{2}^{nm}[i] &=(-1)^m\int_{0}^{\infty}dr\, g_{n2}(r)\sum_{j}^{N}\mathcal{A}_{2}(Z_{j})\mathcal{N}(R_{ij}, \sigma(Z_{j})) H_{m}(r-R_{ij})
    \\
    &=(-1)^m\sum_{j}^{N} \mathcal{A}_{2}(Z_{j}) \int_{0}^{\infty}dr\,g_{n2}(r) \mathcal{N}(R_{ij}, \sigma(Z_{j})) H_{m}(r-R_{ij})
    \label{functional2_shift}
\end{align}
\end{widetext}
Using 
\begin{align}
    f_{m} (r - R_{ij}) = \mathcal{N}(R_{ij}, \sigma(Z_{j})) H_{m}(r-R_{ij})
    \label{fm}
\end{align}
Eq.~\eqref{functional2_shift} can be written as a sum of convolutions
\begin{align}
    P_{2}^{nm}[i]
    =(-1)^m\sum_{j}^{N}~\mathcal{A}_{2}(Z_{j})~(g_{n2} * f_{m})(R_{ij})
\end{align}
\begin{align}
    P_{2}^{nm}[i]
    =(-1)^m\sum_{j}^{N}~\mathcal{A}_{2}(Z_{j})~\mathcal{H}_{2}^{nm}(R_{ij})
    \label{p2nm_final}
\end{align}
and the function $\mathcal{H}_{2}^{nm}$ can be calculated via application of the convolution theorem assuming $g_{n2}(r)$ is square integrable
\begin{align}
    \mathcal{H}_{2}^{nm}(R_{ij}) = \mathcal{F}^{-1}\{\mathcal{F}\{g_{n2}\} \mathcal{F}\{f_{m}\}\}(R_{ij})
    \label{h2nm}
\end{align}
where $\mathcal{F}$ denotes a Fourier transform.
Note that the function $\mathcal{H}_{2}^{nm}$ is unique and independent of the system hence can be evaluated and stored on a pre-defined grid which allows bypassing all the integral evaluations.
\\
In a similar fashion, 3-body functionals can be defined using the 3-body distribution function $\rho_{i}(\theta)$ 
\begin{align}
    \rho_{i}(\theta) &= \braket{{\rho_{i}(\mathbf{r})}|\sum_{jk}f(R_{ij})f(R_{ik})f(R_{jk})\delta(\theta - \theta_{ijk})|{\rho_{i}(\mathbf{r'})}}
\end{align}
\begin{align}
    \rho_{i}(\theta) &\approx \sum_{jk}^{N} \mathcal{A}_{3}(Z_{j},Z_{k})f(R_{ij})f(R_{ik})f(R_{jk})\mathcal{N}(\theta_{ijk}, \sigma(Z_{j},Z_{k}))
    \label{rho3}
\end{align}
where $\theta_{ijk} = \cos^{-1} \left( \frac{\mathbf{R}_{ij} \cdot \mathbf{R}_{ik}}{|\mathbf{R}_{ij}| |\mathbf{R}_{ik}|} \right)
$ is the 3-body inter-atomic angle centered at atom $i$ and $f(R_{ij}) = 1/R_{ij}^{2}$ is chosen such that the Axilrod-Teller-Muto~\cite{axilrod_teller} (ATM) scaling is recovered through the product $f(R_{ij})f(R_{ik})f(R_{jk})$.
This scaling has been successfully applied in other works~\cite{amons_slatm,fchl18}.
Consequently, translationally and rotationally invariant functionals encoding 3-body interactions can be defined in the general form similar to eq.~\eqref{p2nm}
\begin{align}
    P_{3}^{nm}[i] =
    \int_{0}^{\pi}d\theta~g_{n3}(\theta) ~\partial_{\theta}^{m}\rho_{i}(\theta)
\end{align}
Functionals of the derivatives $\partial_{\theta}^{m}\rho_{i}(\theta)$ again allow unique description of the total distribution $\rho_{i}(\theta)$ without generating separate distributions for each unique triplet of chemical elements.
This is common practice with all other $\nu$-body atomic representations which induces a feature vector size scaling of $N_{\rm elem}^{\nu - 1}$ for systems containing $N_{\rm elem}$ number of unique chemical elements.
\\
Using eq.~\eqref{fm}
\begin{align}
   P_{3}^{nm}[i] =(-1)^m\sum_{jk}^{N} \frac{\mathcal{A}_{3}(Z_{j},Z_{k})}{(R_{ij}R_{ik}R_{jk})^2} \int_{0}^{\pi}d\theta g_{n3}(\theta)f_{m}(\theta - \theta_{ijk})
    \label{functional3}
\end{align}
Hence
\begin{align}
    P_{3}^{nm}[i]
    =(-1)^m\sum_{jk}^{N} \frac{\mathcal{A}_{3}(Z_{j},Z_{k})}{(R_{ij}R_{ik}R_{jk})^2}~\mathcal{H}_{3}^{nm}(\theta_{ijk})
    \label{p3nm_final}
\end{align}
where, similar to eq.~\eqref{h2nm}, the function 
\begin{align}
    \mathcal{H}_{3}^{nm}(\theta_{ijk}) = (g_{n3} \circledast f_{m})(\theta_{ijk}) = \mathcal{F}^{-1}\{\mathcal{F}\{g_{n3}\} \mathcal{F}\{f_{m}\}\}(\theta_{ijk})
    \label{h3nm}
\end{align}
 is the circular convolution of the 3-body weighting function $g_{n3}(\theta)$ and the function $f_{m}(\theta - \theta_{ijk})$.
\subsection{Pseudo four-body functionals}
Higher order many-body functionals can be defined in a similar fashion.
For the 4-body terms we define a pseudo 4-body distribution using radial distribution functions for computational efficiency
\begin{align}
    \rho_{i4}(r) \approx \sum_{jkl}^{N} \mathcal{A}_{4}(Z_{j},Z_{k},Z_{l}) \prod_{\{a, b\} \in S_{ijkl}}f_{ab}\mathcal{N}(R_{ab}, \sigma(Z_{b}))
    \label{rho4_1}
\end{align}
with $f_{ab} = \frac{1}{R_{ab}^{2}}$ similar to the 3-body term and $S_{ijkl}=\{a, b\} \in \binom{\{i, j, k, l\}}{2}$ is the set of all 6 unique inter-atomic distances between the 4 atoms $i,j,k,l$.
Eq.~\eqref{rho4_1} can be simplified into a form similar to the 2-body distribution in eq.~\eqref{rho2} as
\begin{align}
    \rho_{i4}(r) = \sum_{jkl}^{N} \mathcal{A}_{4}(Z_{j},Z_{k},Z_{l}) f_{ijkl} \mathcal{N}(R_{ijkl}, \sigma_{ijkl})
    \label{rho4_2}
\end{align}
where $f_{ijkl} = \prod_{\{a, b\} \in S_{ijkl}} \frac{1}{R_{ab}^{2}}$ is the 4-body scaling.
The effective four-body distribution $\mathcal{N}(R_{ijkl}, \sigma_{ijkl}) = \prod_{\{a, b\} \in S_{ijkl}}\mathcal{N}(R_{ab}, \sigma(Z_{b}))$ is obtained via successive application of the Gaussian product theorem
\begin{align}
    \exp\left(-\alpha |r - R_{a}|^2\right) \exp\left(-\beta |r - R_{b}|^2\right) = 
    \notag \\
    \zeta \exp\left(-(\alpha + \beta) |r - R_{p}|^2\right)
\end{align}
where 
\begin{align}
    \zeta = \exp\left(-\frac{\alpha \beta}{\alpha + \beta} |R_a - R_b|^2\right)
\end{align}
\begin{align}
    R_{p} = \frac{\alpha R_{a} + \beta R_B}{\alpha + \beta}
\end{align}
Now using the distribution in eq.~\eqref{rho4_2}, the 4-body functionals are calculated in a similar way to eq.~\eqref{p2nm} 
\begin{align}
    P_{4}^{nm}[i]
    &=\int_{0}^{\infty}dr~g_{n4}(r) ~\partial_{r}^{m}\rho_{i4}(r)
    \notag \\
    &= (-1)^m\sum_{jkl}^{N} \mathcal{A}_{4}(Z_{j},Z_{k},Z_{l}) f_{ijkl}~\mathcal{H}_{4}^{nm}(R_{ijkl})
    \label{p4nm_final}
\end{align}
with 
\begin{align}
    \mathcal{H}_{\nu}^{nm}(t) = \mathcal{F}^{-1}\{\mathcal{F}\{g_{n\nu}\} \mathcal{F}\{f_{m}\}\}(t)
    \label{h4nm}
\end{align}
as before.
Higher order pseudo $\nu$-body functionals can be defined through a similar procedure using radial distributions. 
\subsection{Weighting functions and scaling factors}
Within the presented methodology the choice of the $\nu$-body weighting functions $g_{n\nu}$ is the only hyper-parameter.
We have used very simple weighting functions $g_{n2}(r),g_{n3}(\theta),g_{n4}(r)$ in our work parameterized by an integer $n$.
Two separate types of weighting functions are used for each $\nu$-body case.
For the two and four-body ($\nu=2,4$ respectively) functionals we use simple decaying functions (square-integrable over the positive reals) of the form
\begin{align}
g_{n\nu}(r) =
\left\{
	\begin{array}{ll}
		\exp(-\alpha_{\nu}(n+1)r)\\
        \frac{1}{(r+1)^{(2n+3)}}
 \end{array}
\right.
\label{gn2}
\end{align}
where $\alpha_{\nu} = \alpha_2$, $\alpha_4$ are hyper-parameters (set to 1.5 obtained via a grid-search optimization on 1k random molecules from QM7~\cite{qm7})  of the representation for the two and four-body functionals.
For the 3-body functions we use angular Fourier series terms employed in ACSF based representations~\cite{acsf,fchl19}
\begin{align}
g_{n3}(\theta) =
\left\{
	\begin{array}{ll}
		\cos((2n+1)\theta) - \cos((2n+1)(\theta+\pi))\\
        \sin((2n+1)\theta) - \sin((2n+1)(\theta+\pi))
 \end{array}
\right.
\label{gn3}
\end{align}
\begin{table}[!htb]
\centering
\begin{tabular}{|l|c|c|}
\hline
\textbf{Dataset (size)} & \textbf{\textit{t\textsubscript{gen}} (s)} {\rm cMBDF} & \textbf{\textit{t\textsubscript{gen}} (s)} {\rm MBDF} \\
\hline
QM9~\cite{qm9} (130k)     & 8    & 98   \\
\hline
QM7b~\cite{qm7b} (7.2k)    & 0.1  & 4    \\
\hline
VQM24~\cite{vqm24} (258k)   & 10   & 42   \\
\hline
QMugs~\cite{QMugs} (20k)    & 21   & 251  \\
\hline
\end{tabular}
\caption{Representation generation timing (\textbf{\textit{t\textsubscript{gen}}}, in seconds) comparison between cMBDF and MBDF~\cite{mbdf} representations for 4 different datasets of organic molecules. 
Numbers in brackets of column 1 denote the number of molecules used from the dataset.
}
\label{tab:tgen}
\end{table}
Figure~\ref{fig:convs}B shows the convolved functions $\mathcal{H}_{\nu}^{nm}$ for the $n=1,2$ cases with the weighting functions of the first type from eqs.~\ref{gn2} and ~\ref{gn3}.
We point out here again that once the weighting functions are chosen, these convolutions can be pre-evaluated on a discretized grid as shown in Figure~\ref{fig:convs}B.
For generating the representation vector of an atom within a molecule, only the molecular internal coordinates are required to be calculated.
The functional values $P_{\nu}^{nm}$ can then be obtained by indexing the internal coordinate values on the pre-computed $\mathcal{H}_{\nu}^{nm}$ grid followed by summation. 
The speedup obtained from this is shown in table~\ref{tab:tgen} which tabulates the representation generation timings for organic molecules taken from 4 datasets with different sizes.
These timings are compared to the original MBDF representation which constituted a non-uniform subset of cMBDF in which the functional values were calculated via numerical integration.
In all 4 cases, cMBDF is faster by an order of magnitude or more despite evaluating 7$\times$ (40 for cMBDF vs 5 for MBDF) more functionals per atom.
Representation generation timings for some other commonly used representations (also used in this study below) can be found in ref.~\cite{mbdf} for the same molecules.
\\
The scaling pre-factors $\mathcal{A}_{\nu}(Z_{1},Z_{2}..Z_{\nu-1})$ used alongside all distributions (eqs.~\ref{rho2},\ref{rho3},\ref{rho4_2}) are calculated as geometric mean of a function encoding the chemical identity of the elements
\begin{align}
    \mathcal{A}(Z) = \log(P+1)G
    \label{alc_scale}
\end{align}
\begin{align}
    \mathcal{A}_{\nu}(Z_{1},Z_{2}..Z_{\nu-1}) = (\prod_{j=1}^{\nu-1} \mathcal{A}(Z_{j}))^\frac{1}{\nu-1}
\end{align}
where $P$, $G$ denote the period and group number of the chemical element $Z$ in the periodic table.
The form chosen in eq.~\eqref{alc_scale} ensures scaling factors for elements from the same group are more similar than from the same period.
Similarly, the standard deviations $\mathcal{\sigma}(Z_{1},Z_{2}..Z_{\nu-1})$ used alongside all distributions (eqs.~\ref{rho2},\ref{rho3},\ref{rho4_2}) are calculated as weighted means of the van-der-Waals (vdW) radii ($r_{\rm vdW}$) of each chemical element $Z$ involved
\begin{align}
    \mathcal{\sigma}(Z_{1},Z_{2}..Z_{\nu-1}) = \frac{\sum_{j=1}^{\nu-1}Z_{j} ~r_{\rm vdW}(Z_{j})}{\sum_{j=1}^{\nu-1}Z_{j}}
\end{align}
\\
The functional values evaluated from eqs.~\ref{p2nm_final},\ref{p3nm_final},\ref{p4nm_final} are then concatenated to form the feature vector $\mathbf{P}_{i}$ describing atom $i$
\begin{align}
    \mathbf{P}_{i}
     = [P_{2}^{00}[i]...P_{2}^{nm}[i],P_{3}^{00}[i]...P_{3}^{nm}[i],P_{4}^{00}[i]....P_{4}^{nm}[i]]
\end{align}
The entire procedure is summarized as a schematic in Figure~\ref{fig:convs}A.
\\
This leads to a class of compact and systematically improvable atomic descriptors as controlled by the three integers $\nu$ (many body-order), $m$ (derivative-order) and $n$ (weighting function-order).
The size (dimensionality) of the atomic feature vector is given by the product $2(\nu-1) (m+1) n$ ($\mathbf{P}_{i}~\epsilon ~\mathbb{R}^{2(\nu-1) (m+1) n}$) and is invariant to the chemical species, system size and cut-offs employed.
Figure~\ref{fig:errors_gs} shows the predictive performance variation of cMBDF with the three integers $\nu$, $m$ and $n$.
As expected, the many-body order $\nu$ has the largest effect on the accuracy.
However, a similar performance as the four-body representation can be reached by raising $m$ and $n$ to larger values while keeping $\nu$=3.
This would be beneficial for larger systems where four-body and higher order functionals can become expensive to evaluate.
Nevertheless, four-body terms are known to be crucial in some cases~\cite{incompleteness_ceriotti} and increase the representation sensitivity~\cite{parsaeifard2021assessment}.
The 4-body terms can also be important for some physical properties other than energetics as shown in the results.
\\
Throughout the rest of the work we have used $m=4$, $n=2$ along with $\nu=3$ (denoted cMBDF) or $\nu=4$ (denoted cMBDF (4-body)) leading to atomic feature vector lengths of 40 and 60 respectively.
\subsection{Gradients for responses}
Gradients required for calculating response properties can be evaluated efficiently as well.
We first note that the gradients of the $\mathcal{H}_{\nu}^{nm}$ convolved functions with respect to the nuclear position $\mathbf{R}_{a}$ can be evaluated easily via application of the chain rule 
\begin{align}
    \nabla_{\mathbf{R}_{a}} \mathcal{H}_{\nu}^{nm}(x(\mathbf{R}_{a})) = \partial_{x} \mathcal{H}_{\nu}^{nm}(x) \nabla_{\mathbf{R}_{a}}x(\mathbf{R}_{a})
\end{align}
and
\begin{align}
    \partial_{x} \mathcal{H}_{\nu}^{nm}(x) = \mathcal{F}^{-1}\{\mathcal{F}\{g_{n \nu}\} \mathcal{F}\{\partial_{x}f_{m}\}\}(x) = -\mathcal{H}_{\nu}^{n(m+1)}(x)
\end{align}
where the first equality follows from Leibniz rule and we have used the relation $\partial_{x} f_{m} = -f_{m+1}$ from eq.~\eqref{fm} for the second equality. 
Hence, the derivative on $\mathcal{H}_{\nu}^{nm}$ essentially raises the Hermite polynomial degree by 1.
Consequently
\begin{align}
    \nabla_{\mathbf{R}_{a}} \mathcal{H}_{\nu}^{nm}(x(\mathbf{R}_{a})) = -\mathcal{H}_{\nu}^{n(m+1)}(x)\nabla_{\mathbf{R}_{a}}x(\mathbf{R}_{a})
    \label{conv_grad}
\end{align}
Note that due to the symmetry of convolutions the following expression also holds
\begin{align}
    \frac{\partial \mathcal{H}_{\nu}^{nm}(x)}{\partial x} = \mathcal{F}^{-1}\{\mathcal{F}\{\partial_{x}g_{n \nu}\} \mathcal{F}\{f_{m}\}\}(x)
\end{align}
however a similar relation to $\partial_{x} f_{m} = -f_{m+1}$ does not hold for the weighting function derivatives $\partial_{x}g_{n \nu}$ in general and is not used in our work.
\\
Using eq.~\eqref{conv_grad} the gradients of the functionals $P_{\nu}^{nm}$ can be readily evaluated as
\begin{widetext}
\begin{align}
\nabla_{\mathbf{R}_{a}}P_{2}^{nm} = (-1)^{m+1}\sum_{j}^{N}~\mathcal{A}_{2}(Z_{j})~\mathcal{H}_{2}^{n(m+1)}(R_{ij}) \nabla_{\mathbf{R}_{a}}R_{ij}
\end{align}
\begin{align}
\nabla_{\mathbf{R}_{a}}P_{3}^{nm} = (-1)^{m}\sum_{jk}^{N}~\mathcal{A}_{3}(Z_{j},Z_{k})\left[\mathcal{H}_{3}^{n(m)}(R_{ij}) \nabla_{\mathbf{R}_{a}}f_{ijk} - f_{ijk}\mathcal{H}_{3}^{n(m+1)}(R_{ij}) \nabla_{\mathbf{R}_{a}}\theta_{ijk}\right]
\end{align}
\begin{align}
\nabla_{\mathbf{R}_{a}}P_{4}^{nm} = (-1)^{m}\sum_{jkl}^{N} \mathcal{A}_{4}(Z_{j},Z_{k},Z_{l}) \left[\mathcal{H}_{4}^{nm}(R_{ijkl})\nabla_{\mathbf{R}_{a}}f_{ijkl} - f_{ijkl}\mathcal{H}_{4}^{n(m+1)}(R_{ijkl})
    \nabla_{\mathbf{R}_{a}}R_{ijkl}\right]
\end{align}
\end{widetext}

\begin{figure*}[!htb]
    \centering
    \includegraphics[width=\linewidth]{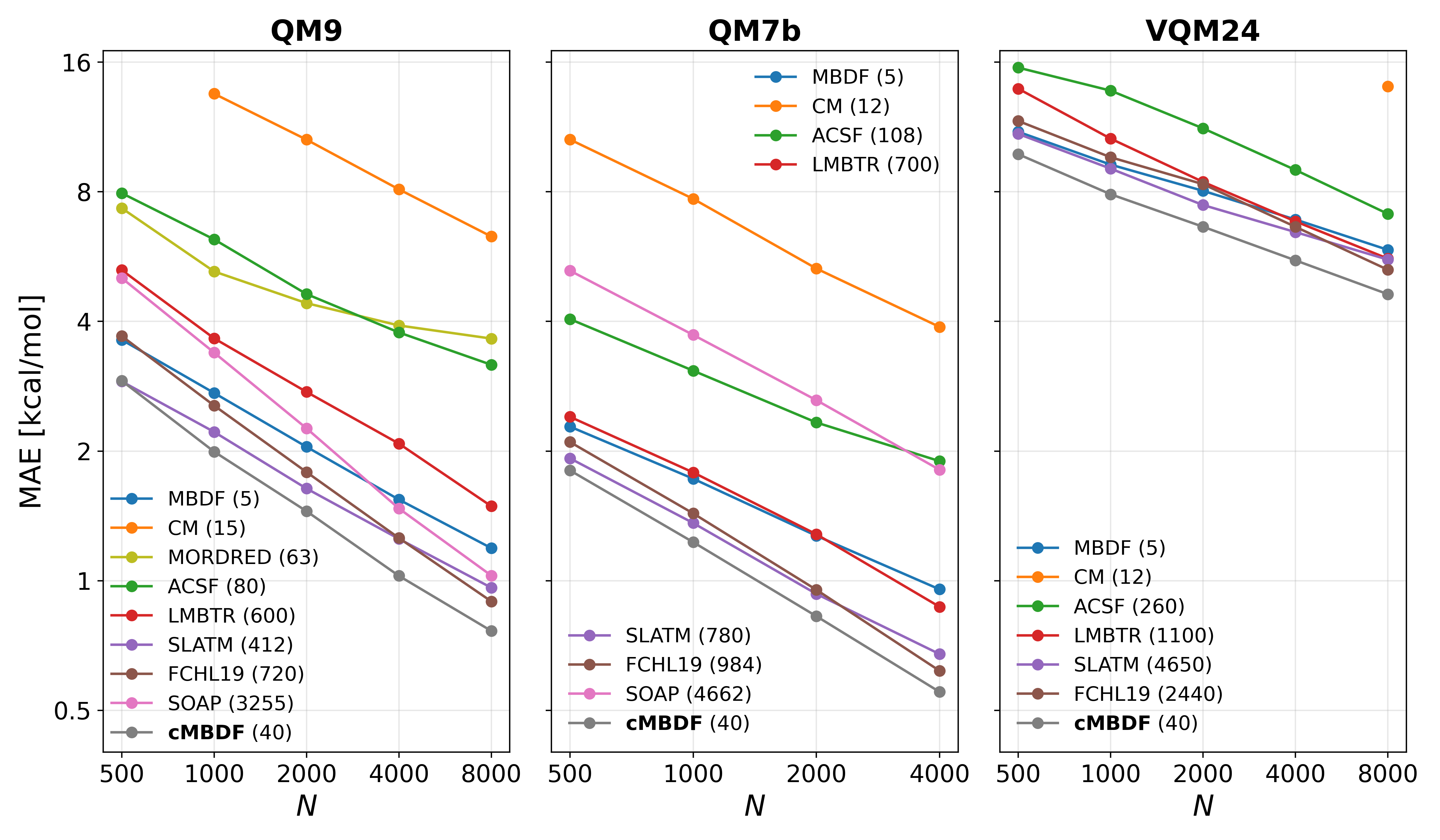}
    \caption{Kernel ridge regression (KRR) based learning curves (model prediction error as a function of training set size) for molecular atomization energies from the QM9~\cite{qm9}, QM7b~\cite{qm7b,qm7} and VQM24~\cite{vqm24} datasets of small organic and inorganic molecules.
    Comparison to some commonly used representations alongside KRR models is shown across all 3 datasets.  
    Numbers in legend denote size (dimensionality) of the atomic feature vector mapping induced by the corresponding representation.
    For the molecular representations (CM, SLATM, MORDRED) the dimensionality noted is the molecular feature vector size divided by the number of atoms.
    } \label{fig:lc_energies}
\end{figure*}
\begin{figure}[!htb]
    \centering
    \includegraphics[width=\linewidth]{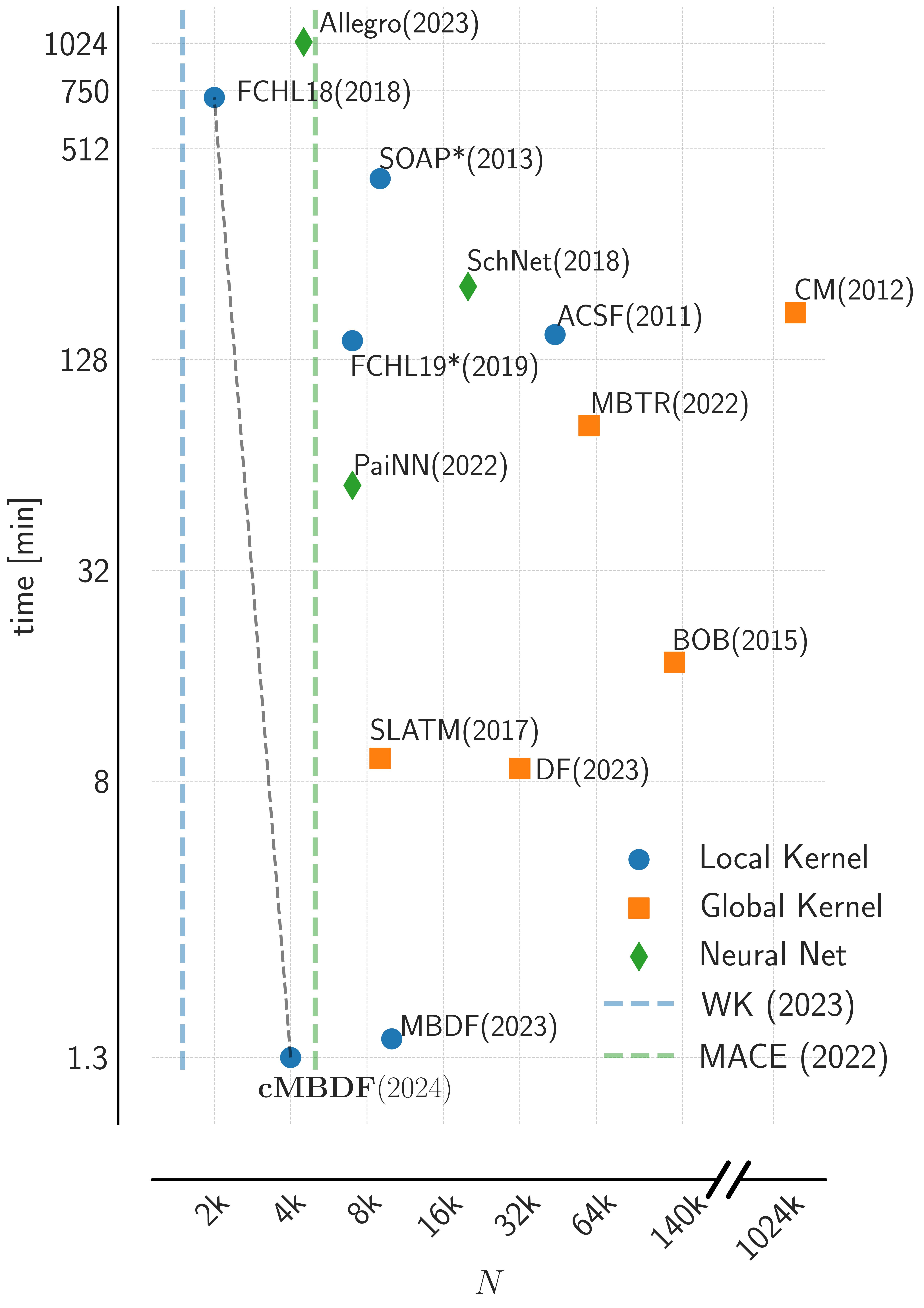}
    \caption{Plot showing trade-off between computational and data efficiency for various ML methods.
    X-axis denotes the (minimum) number of training samples required to achieve chemically accurate (MAE = 1 kcal/mol) atomization energy predictions on the entire QM9 dataset. 
    Y-axis plots the model training and prediction timing for the same task.
    Data for models other than cMBDF, Wigner Kernels (WK)~\cite{wigner_kernels} and MACE~\cite{mace} is taken from ref.~\cite{mbdf}.
    The WK and MACE lines denote the (minimum) number of training samples required for chemically accurate predictions on the QM9 dataset taken from refs.~\cite{wigner_kernels},  ~\cite{mace_qm9} respectively.}
    \label{fig:pareto}
\end{figure}
\begin{figure*}[!htb]
    \centering
    \includegraphics[width=\linewidth]{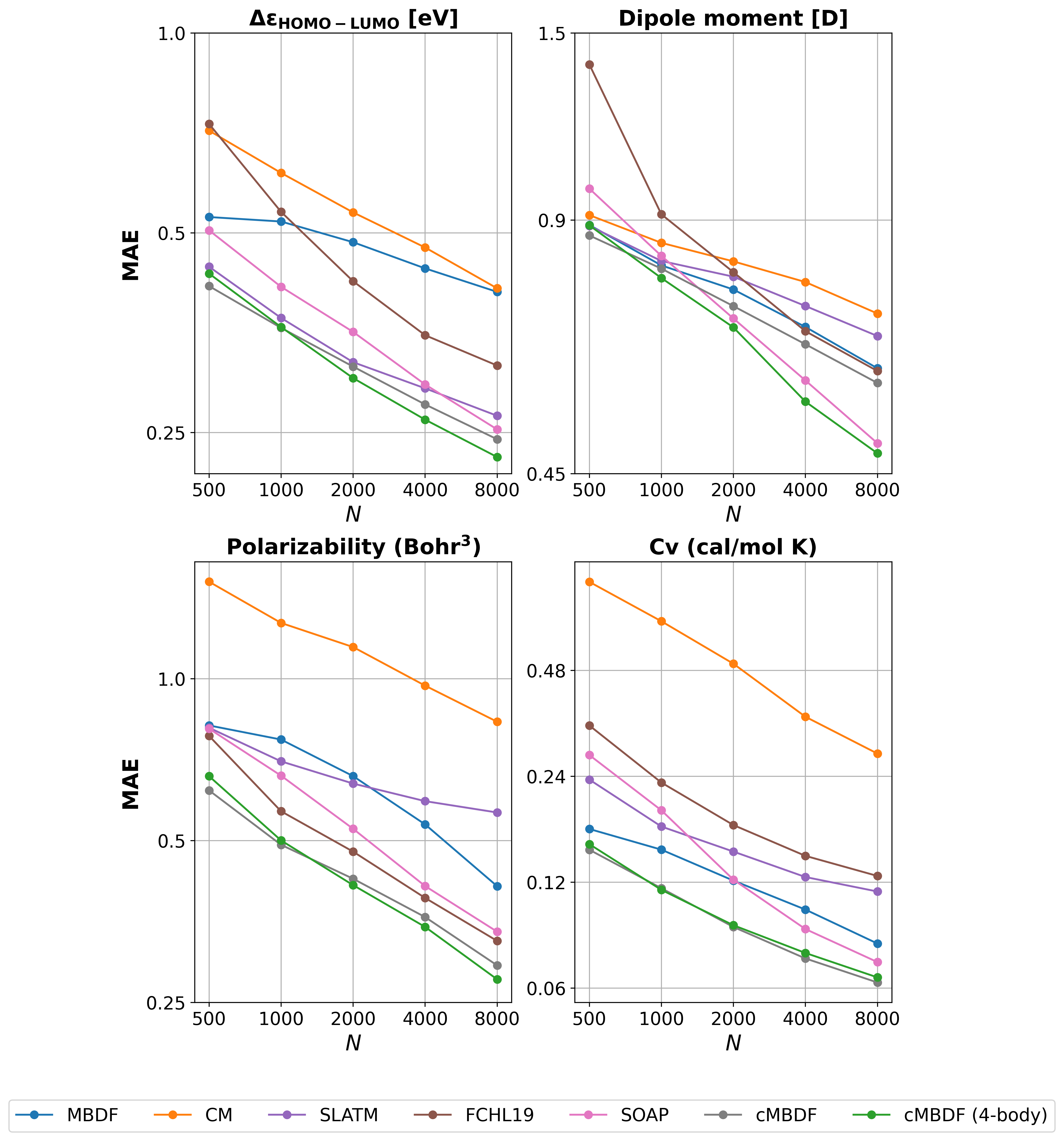}
    \caption{Learning curves for some intensive molecular properties from the QM9~\cite{qm9} dataset.
    Data points for learning curves other than cMBDF and cMBDF (4-body) are taken from ref.~\cite{mbdf}.
    } \label{fig:lc_others}
\end{figure*}
\begin{figure*}[!htb]
    \centering
    \includegraphics[width=\linewidth]{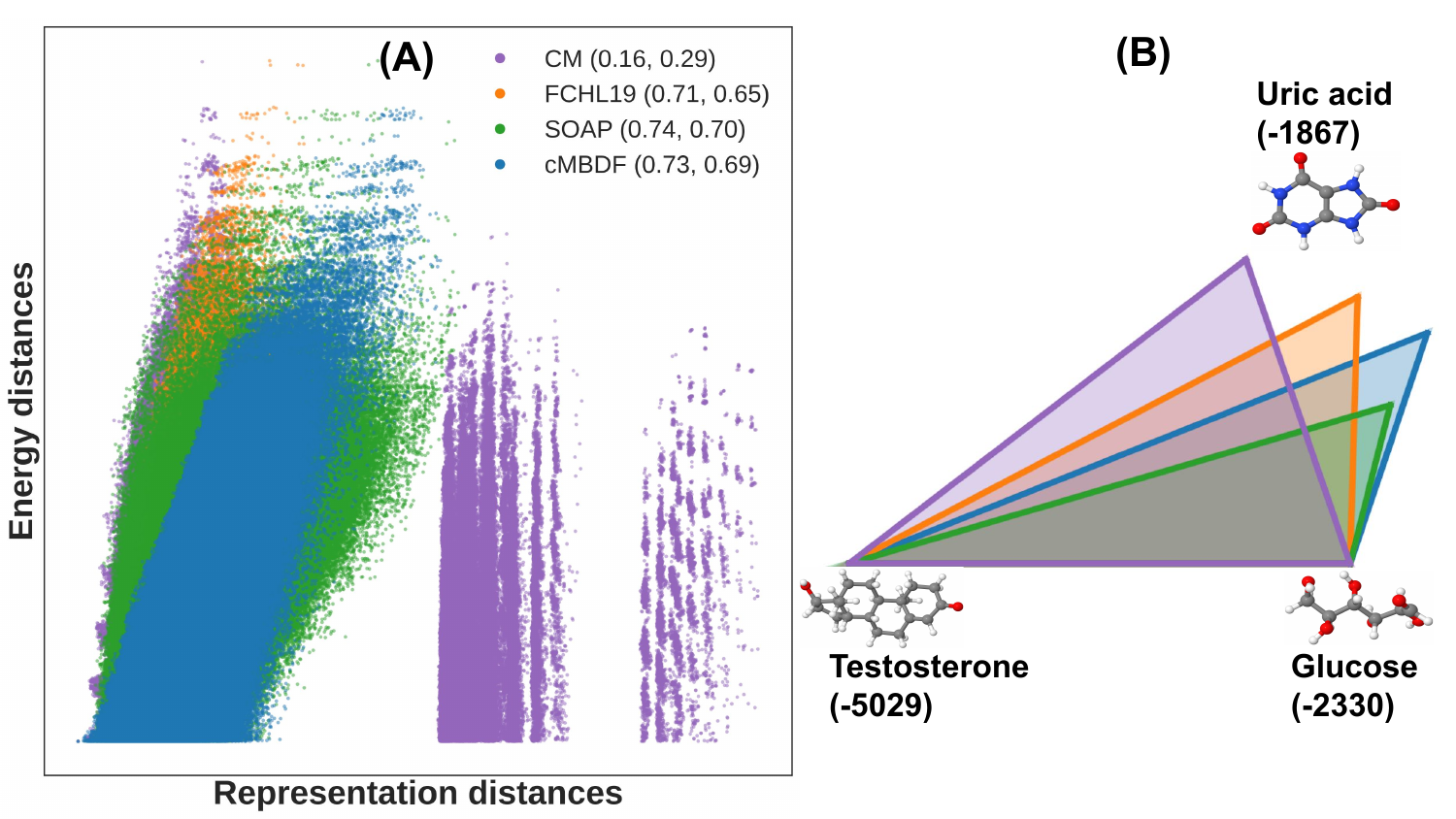}
    \caption{Correlation analysis between energetic and representation distances. (A) Correlation plot between atomization energy and molecular representation distances (Frobenius norm) for all molecules from the QM7b~\cite{qm7b} dataset. 
    All differences were normalized by the mean distance over the dataset.
    Numbers in the legend denote the Pearson and Spearman's rank correlation copefficient values respectively.
    (B) Molecular representation matrix (relative) distances (Frobenius norm) between 3 biologically relevant molecules. 
    Sides of the triangles are proportional to the representation distance between the two molecules at the vertices. 
    Distances shown are ratios to the  Testosterone-Glucose distance for all 4 representations.
    Numbers in brackets below molecule name denote PBE0/def2-QZVP~\cite{pbe0,def2tzvp} calculated atomization energies (using \texttt{PySCF}~\cite{sun2018pyscf}) in kcal/mol.
    } 
    \label{fig:special_mols}
\end{figure*}
a
where $\nabla_{\mathbf{R}_{a}}f_{ijk} = \nabla_{\mathbf{R}_{a}}\frac{1}{(R_{ij}R_{ik}R_{jk})^2}$ and $\nabla_{\mathbf{R}_{a}} f_{ijkl} = \nabla_{\mathbf{R}_{a}}\prod_{\{a, b\} \in S_{ijkl}} \frac{1}{R_{ab}^{2}}$ are straightforward to evaluate using the internal coordinate gradients $\nabla_{\mathbf{R}_{a}} R_{ij}$ and $\nabla_{\mathbf{R}_{a}} \theta_{ijk}$
\begin{align}
    \nabla_{\mathbf{R}_{a}} R_{ij}
     = &(\delta_{aj} - \delta_{ai})\frac{\mathbf{R}_{j} - \mathbf{R}_{i}}{R_{ij}} 
     \\
    \nabla_{\mathbf{R}_{a}} \theta_{ijk}
     = &\frac{\cos \theta_{ijk}\nabla_{\mathbf{R}_{a}} R_{ij}}{R_{ij}|\sin{\theta_{ijk}}|} - \frac{\nabla_{\mathbf{R}_{a}} R_{ik}}{|\sin{\theta_{ijk}}|R_{ij}}
\end{align}

\section{Results and discussions}
\subsection{Data efficiency}
We begin our discussion with the classic benchmark of learning atomization energies from a few datasets of small organic and inorganic molecules.
ML model details can be found in the Kernel based methods section of SI.
Figure~\ref{fig:lc_energies} shows learning curves (ML model prediction error as a function of training set size) of atomization energies from the QM9~\cite{qm9}, QM7b~\cite{qm7b} and VQM24~\cite{vqm24} datasets.
For comparison we also plot some of the most commonly employed atomic (atom centred symmetry functions (ACSF)~\cite{acsf}, local many-body tensor representation (LMBTR)~\cite{mbtr}, Faber-Christensen-Huang-Lilienfeld 2019 variant (FCHL19)~\cite{fchl19}, smooth overlap of atomic positions (SOAP)~\cite{soap}), molecular (Coulomb matrix (CM)~\cite{CM}, spectrum of London and Axilrod-Teller-Muto (SLATM)~\cite{amons_slatm}) and graph-based (MORDRED~\cite{mordred}) representations alongside kernel based ML models.
We restrict our analysis to upto $\sim$10,000 training data points which constitutes the low-training data regime where kernel methods are more efficient than their deep-learning counterparts~\cite{gary2023,felix_google}.
Figure~\ref{fig:lc_energies} also shows the size (dimensionality) of the feature vector mapping (per atom) induced by each representation in the legend.
Across all three datasets, a similar ordering in terms of both the accuracy and representation size is seen for all of the tested representations.
For representations other than cMBDF, the accuracy is well correlated with the representation size which, however, has a strong impact on the computational cost of each method (see Figure~\ref{fig:pareto}).
cMBDF stands out as an exception as it remains the most accurate (data efficient) while retaining compactness (computational efficiency).
While being comparable in size to the CM representation across all three datasets, cMBDF requires nearly 16$\times$ less training data to reach the same accuracies.
Furthermore, while the atomic feature vector size scales with either the number of atoms or unique chemical species for all other representations, cMBDF remains constant size at 40 dimensions.
The effect of this is especially pronounced on the more chemically diverse VQM24~\cite{vqm24} dataset which contains 10 unique chemical elements.
As can be seen from figure~\ref{fig:lc_energies}, cMBDF remains nearly two orders of magnitude more compact than the other best performing representations while showing a lower predictive error in the low training data regime.
\\
\subsection{Compute efficiency}
This compactness directly translates to computational efficiency.
On the VQM24 dataset, for instance, generation of the entire learning curves (training, optimization and predictions) in figure~\ref{fig:lc_energies} with the FCHL19 vs cMBDF representations required $\textbf{23 hours vs 8 minutes}$ respectively (on a compute node with 36 core 4.8GHz Intel Xeon W9-3475X/1 TB DDR5 ECC RAM).
This is further demonstrated in figure~\ref{fig:pareto} which shows the model training and prediction time vs training data requirement to reach chemically accuracy (MAE $<$ 1 kcal/mol) on the entire QM9~\cite{qm9} dataset.
The dashed grey line indicates the optimal Pareto front for computational vs data efficiency tradeoff. 
cMBDF significantly improves upon other representations when employed alongside kernel based methods as it shows the fastest model timings while being the third most data-efficient method (after Wigner Kernels~\cite{wigner_kernels} and FCHL18~\cite{fchl18}).
cMBDF reaches chemical accuracy on the entire QM9 dataset after training on only 4,000 molecules and required 1.3 minutes (for training and subsequent prediction on 100k QM9 molecules) of compute time.
In comparison to the more data-efficient FCHL18~\cite{fchl18} representation, cMBDF is $\sim$550$\times$ faster.
\subsection{Other quantum properties}
This computational efficiency makes cMBDF ideal for the training and testing of new models across various regions of chemical space, as well as for various tasks.
Learning capacity for properties other than energetics are demonstrated in figure~\ref{fig:lc_others}.
Evidently, cMBDF also retains its good data efficiency for other physical properties of interest.
Furthermore, the performance seems transferable to intensive properties such as HOMO-LUMO gaps and electrostatic moments which are not amenable to atomic partitioning schemes.
This has been noted in other work with various methods proposed to deal with the learning of HOMO-LUMO gaps~\cite{bernard_HOMO_LUMO,chen2023physics,ganna_homo}.
Similarly, methods have been proposed to improve the performance of ML models for the learning of dipole moments such as the inclusion of response terms in the loss function~\cite{OQML}.
Appreciable improvements can be observed through inclusion of 4-body terms (especially for dipole moments) for these properties in cMBDF.
Furthermore, due to the linear scaling of cMBDF atomic vector sizes with the many-body order, the computational cost is negligibly affected upon inclusion of the 4-body term (atomic feature vector size changes from 40 to 60).
Due to this versatility and computational efficiency, cMBDF has been successfully applied to adaptive-ML based methods for learning optimal exact-exchange mixing fractions with the PBE0~\cite{pbe0} functional~\cite{apbe0} and optimal scaling factors for Pople-type Gaussian basis sets~\cite{abasis}.
Learning curves for both tasks can be found in the SI (and the corresponding studies) and show a similar trend.
\\ 
\subsection{Sensitivity analysis}
To analyze the effectiveness of cMBDF, we examine the relation of representation and energetic differences between various molecules.
Figure~\ref{fig:special_mols}A shows a correlation plot between atomization energy and representation matrix distances between all pairs of molecules from the QM7b~\cite{qm7b} dataset.
For comparison we also plot 3 other commonly used representations across chemical compound space.
To measure the linear and non-linear correlations, we calculated the Pearson and Spearman's rank correlation coefficient values for the 4 representations tested.
A good linear correlation between the energetic and representation distances can be observed for the FCHL19, SOAP and cMBDF molecular representations.
This is important since kernel based methods operate on the hypothesis that similar systems are expected to have similar labels.
The correlation, however, can be non-linear due to the mapping induced by the kernel function.
The Spearman's rank correlation coefficient takes this into account as it measures the degree of monotonicity between the two variables.
For both coefficients, cMBDF shows strong correlations similar to the SOAP representation which induces a feature vector mapping 2 orders of magnitude larger than cMBDF (4662 vs 40 dimensions).
This likely underpins the greater accuracy achieved by cMBDF based kernel models in the low training data regime (fig.~\ref{fig:lc_energies}B).
\\
This correlation can also be seen for larger systems across chemical compound space.
Figure~\ref{fig:special_mols}B shows representation distance between 3 biologically relevant molecules of different size and composition.
Sides of the triangles in the figure are proportional to the relative distances between representation matrices of the molecules at the vertices.

Evidently, the relative distances between molecular representations induced by cMBDF align well with the DFT calculated energetic differences of the 3 molecules.
The testosterone-uric acid relative distance is the largest with cMBDF between the 4 representations even with cMBDF being significantly more compact than the other 3 representations.
The smaller uric acid-glucose distance with cMBDF also aligns well with the relative energetic difference which is the smallest amongst the the 3 pairs.
This suggests that, despite their compact size, cMBDF feature vectors efficiently capture rich structural and compositional information, resulting in its strong performance on learning tasks.
\section{Conclusions}
In this work we have introduced a systematically improvable class of compact atomic representations for use throughout chemical compound space.
Convolutional many-body distribution functionals (cMBDF) encode the chemical environment of an atom through a set of translationally and rotationally invariant functionals of the atomic density.
The functional values are efficiently evaluated via fast Fourier transforms using the convolution theorem which can be evaluated and stored on a pre-defined grid.
Weighting functions of various types can be incorporated to capture physical interactions (short and long range) of interest efficiently via the atomic density functionals.
The atomic feature vector size is invariant to the cut-offs employed, system size and composition leading to a compact and constant size atomic descriptor.
cMBDF is shown to outperform other commonly used representations for the learning of a variety of physical properties across chemical space while remaining nearly two orders of magnitude more compact.
Due to its computational efficiency and versatility, cMBDF can lead to significantly faster prototyping, training and testing of quantum machine learning models for a variety of tasks across chemical compound space.
Future work will include a study of cMBDF gradients and their applicability across geometrical changes for relaxing geometries, identifying transition states~\cite{heinen_transition} and molecular dynamics.
\section*{Data and code}
Python implementation for generating cMBDF representations along with gradients is openly available at
https://github.com/dkhan42/cMBDF
\\
It relies on the \texttt{Numpy}~\cite{numpy}, \texttt{Scipy}~\cite{2020SciPy-NMeth} and \texttt{Numba}~\cite{numba} python libraries.
\\
\texttt{QMLcode}~\cite{qmlcode2017} library was used for implementing KRR models and generating the CM, SLATM, FCHL19 representations with default parameters.
\\
\texttt{Dscribe}~\cite{dscribe,dscribe2} library was used to generate the ACSF, LMBTR, SOAP representations with default parameters.
\begin{acknowledgments} 
D.K. is thankful for discussions with S. L. Krug and O. Trottier.
We acknowledge the support of the Natural Sciences and Engineering Research Council of Canada (NSERC), RGPIN-2023-04853.
O.A.v.L. has received funding from the European Research Council (ERC) under the European Union’s Horizon 2020 research and innovation programme (grant agreement No. 772834).
This research was undertaken thanks in part to funding provided to the University of Toronto's Acceleration Consortium from the Canada First Research Excellence Fund,
grant number: CFREF-2022-00042.
O.A.v.L. has received support as the Ed Clark Chair of Advanced Materials and as a Canada CIFAR AI Chair.  
\end{acknowledgments}
\section*{Supplementary Information: Convolutional many body distribution functional representations}
\renewcommand{\thefigure}{S\arabic{figure}}
\begin{figure}[!htbp]
    \centering 
    \includegraphics[width=\columnwidth]
    {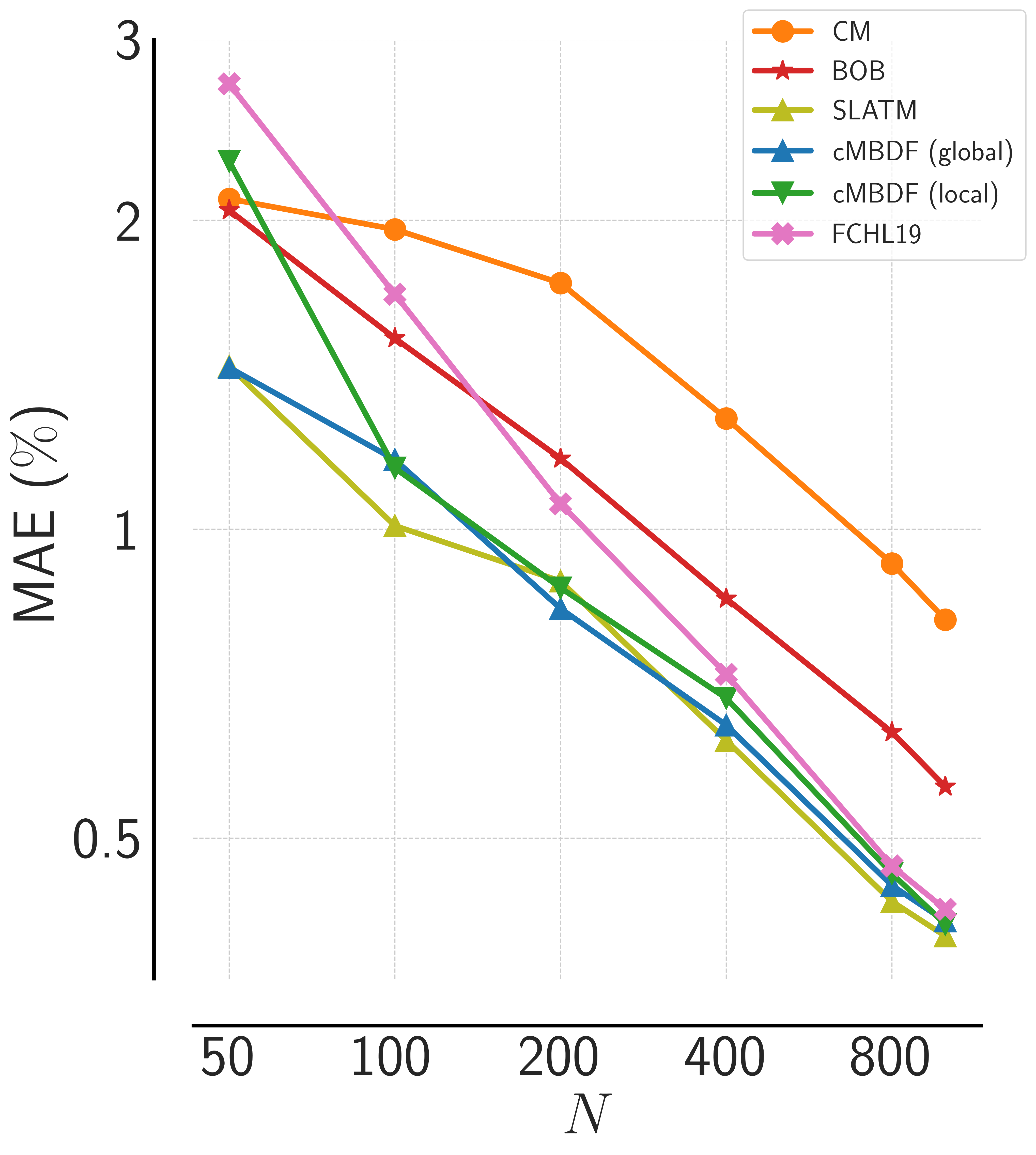}
    \caption{Learning curves showing prediction error for optimal HF admixture ratio ($a_{\mathrm{opt}}$) with the PBE0 functional~\cite{pbe0} as a function of training set size for the representations Coulomb Matrix (CM)\cite{CM_and_qm7}, Bag of Bonds (BOB)\cite{bob}, Spectrum of London and Axilrod-Teller-Muto potentials (SLATM)\cite{amons}, Faber-Christensen-Huang-Lilienfeld 19 (FCHL19)\cite{fchl19} and convolutional Many Body Distribution Functionals (cMBDF)\cite{mbdf}.
Training and testing (200 out-of-sample amons) is performed on QM5 dataset~\cite{bing_DMC}.
$a_{\mathrm{opt}}$ values were calculated by optimizing the aPBE0 atomization energy to CCSD(T) atomization energy for each system.
Figure taken from ref.~\cite{apbe0}.
    }
    \label{fig:learning_curves_amons}
\end{figure}
\begin{figure*}[!htbp]
          \centering           
          \includegraphics[width=\linewidth]{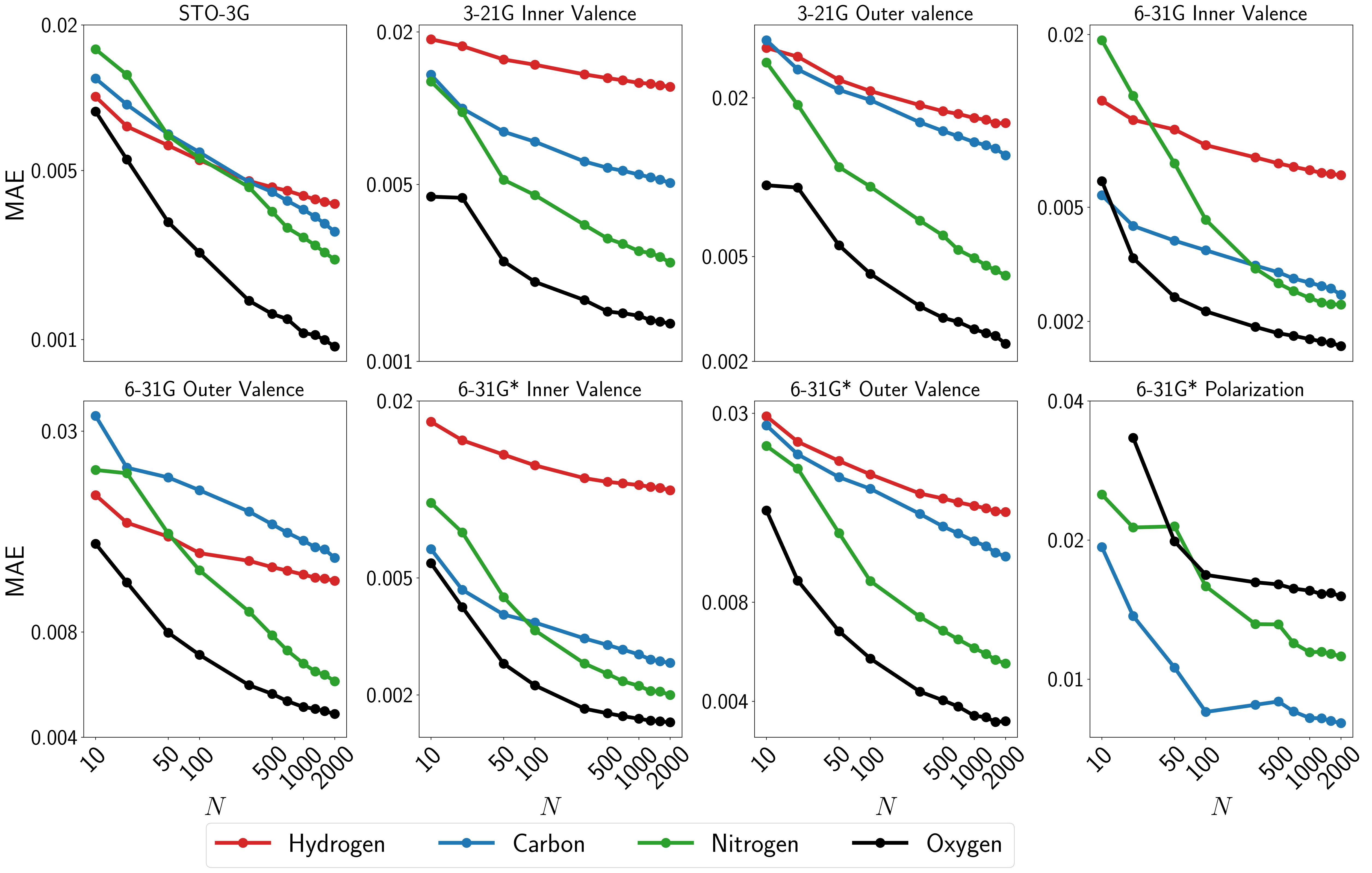}
          \caption{Mean absolute errors (MAE) of predicted optimal scaling factors for STO-3G (valence orbitals), 3-21G, 6-31G (both inner and outer valence) and 6-31G* (inner and outer valence and polarization functions) as a function of training set size (number of molecules) using cMBDF on a validation set of 500 out-of-sample QM9 molecules.
          Figure taken from ref.~\cite{abasis}.
          }
     \label{fig:lc_zeta}
 \end{figure*}
\section{Kernel based methods}
Unless specified otherwise, the ML model used throughout this work with all representations including cMBDF is the  established Kernel Ridge Regression~\cite{vapnik1999nature} (KRR), as also widely adopted by authors in the Lecture Notes in Physics book on quantum machine learning~\cite{MachineLearningMeetsQuantumPhysics2020book}. 
The reason for this choice is primarily its excellent performance in the low training data regime~\cite{felix_google, gary2023} and ease of usage.
Briefly, in KRR the prediction $y_{q}$ for a query system is obtained as a weighted sum of similarity measures to all atoms/molecules in the training set
\begin{equation}
    y_{q} = \sum_{J}^{N_{\mathrm{train}}} \alpha_{J}k(\mathbf{X}_{q}, \mathbf{X}_{J})
\end{equation}
where $\alpha_{j}$ are the regression weights, $\mathbf{X}$ are molecular representation matrices or atomic representation vectors, and $k(.,.)$ denotes a kernel function acting as a similarity measure.
The regression weights $\mathbf{\alpha}$ are obtained from the set of training labels $\mathbf{y}^{\mathrm{train}}$ via the following equation
\begin{equation}
    \mathbf{\alpha} = (\mathbf{K} + \lambda \cdot \mathbf{I})^{-1} \mathbf{y}^{\mathrm{train}}
    \label{eq:solution_alpha}
\end{equation}
where $\mathbf{K}$ is the kernel matrix of the training set, $\lambda$ is a regularization parameter and $\mathbf{I}$ is the identity matrix.
The kernel function primarily used in our work is the screened atomic Gaussian kernel
\begin{equation}
    k(\mathbf{X}_{I}, \mathbf{X}_{J}) = \sum_{\mu\epsilon I}\sum_{\nu\epsilon J}\mathbf{\delta}_{Z_{\mu},Z_{\nu}}  \exp\left( -\frac{\vert \vert \mathbf{P}_{I\mu} - \mathbf{P}_{J\nu} \vert \vert^{2}_{2}}{2 l^2} \right)~\label{eq:local_gaussian}
\end{equation}
where $\mathbf{P}_{I\mu}$ denotes the representation vector of atom $\mu$ within molecule $I$, $l$ denotes the length-scale hyper-parameter of the kernel and $\mathbf{\delta}_{Z_{\mu},Z_{\nu}}$ denotes a Kronecker Delta over the nuclear charges $Z_{\mu},Z_{\nu}$ which restricts the similarity measurement between atoms of the same chemical element\cite{fchl19}.
The form of the kernel function in eq.~\ref{eq:local_gaussian} partitions the system into atomic contributions and measures similarities between each atomic environment in the training set and query system.
The partitioning also leads to atom-index invariance due to the summation.
\\
Alternatively, feature vector mappings describing the entire system can be used resulting in molecular (or global) kernels of the form
\begin{align}
    k(\mathbf{X}_I, \mathbf{X}_J) = \exp{\left( -\frac{\vert \vert \mathbf{X}_I - \mathbf{X}_J \vert \vert^{2}_{2}}{2 l^2} \right)}
    \label{eq:kernel_gaussian}
\end{align}
where $\mathbf{X}_I$ now denotes a feature vector mapping of the molecule $I$.
This form is usually more amenable to the learning of intensive properties that cannot be effectively partitioned onto atomic contributions~\cite{amons_slatm,mbdf}.
\\
To achieve atom-index invariance in the global representation form $\mathbf{X}_I$ with cMBDF, we employ "bagging" in a similar fashion as the bag-of-bonds (BOB)~\cite{bob} representation.
Each set of functionals $\mathbf{P}[i]$ describing atom $i$ is first sorted, followed by arrangement in a pre-specified order based on the chemical identities of the atoms in the molecule. 
This restricts the distance measurement in eq.~\ref{eq:kernel_gaussian} to atomic species of the same type analogous to the Kronecker Delta in eq.~\ref{eq:local_gaussian}. 
While this method leads to siginificant gain in accuracy compared to a simple sorting~\cite{bob}, it introduces a scaling of the cMBDF molecular vector $\mathbf{X}_I$ dependent on the unique chemical species within the dataset.
\\
The length-scale ($l$) kernel hyper-parameter in eqs.~\ref{eq:local_gaussian}, ~\ref{eq:kernel_gaussian} and the regularizer ($\lambda$) in eq.~\ref{eq:solution_alpha} were optimized via grid-search.
For the local kernels we used logarithmic grids of $\left[0.1(2^{n}) ~ \forall n \epsilon \{0,14\}\right]$ for $l$ and $\left[10^{-3n}~ \forall n \epsilon \{1,4\} \right]$ for $\lambda$.
For global kernels we use the grid $\left[10^{n} ~ \forall n \epsilon \{2,8\} \right]$ for $l$.
\bibliographystyle{apsrev4-1}
\bibliography{literature.bib}
\end{document}